\documentclass{bvm2022} 

\begin{document}

\selectlanguage{english} 

\title{Deep Learning-Based Automatic Assessment of AgNOR-scores in Histopathology Images}

\subtitle{}

\titlerunning{Automatic Assessment of AgNOR-scores}

\author{
	Jonathan \lname{Ganz} \inst{1}, 
	Karoline \lname{Lipnik} \inst{2}, 
	Jonas \lname{Ammeling} \inst{1},
        Barbara \lname{Richter} \inst{2},
        Chloé \lname{Puget} \inst{3},
        Eda \lname{Parlak} \inst{2},
        Laura \lname{Diehl} \inst{3},
        Robert \lname{Klopfleisch} \inst{3},
        Taryn~A. \lname{Donovan} \inst{4},
        Matti \lname{Kiupel} \inst{5},
        Christof~A. \lname{Bertram} \inst{2}, 
	Katharina \lname{Breininger} \inst{6} and
	Marc \lname{Aubreville} \inst{1}
}

\authorrunning{Ganz et al.}

\institute{
\inst{1} Technische Hochschule Ingolstadt, Ingolstadt, Germany\\
\inst{2} Institute of Pathology, University of Veterinary Medicine Vienna, Vienna, Austria\\
\inst{3} Institute of Veterinary Pathology, Freie Universität Berlin, Berlin, Germany\\
\inst{4} The Schwarzmann Animal Medical Center, New York, USA\\
\inst{5} Department of Pathology and Diagnostic Investigation, Michigan State University, East Lansing, USA\\ 
\inst{6} Department Artificial Intelligence in Biomedical Engineering, Friedrich-Alexander-Universität Erlangen-Nürnberg, Erlangen, Germany}

\email{jonathan.ganz@thi.de}
\maketitle

\begin{abstract}
Nucleolar organizer regions (NORs) are parts of the DNA that are involved in RNA transcription. Due to the silver affinity of associated proteins, argyrophilic NORs (AgNORs) can be visualized using silver-based staining. The average number of AgNORs per nucleus has been shown to be a prognostic factor for predicting the outcome of many tumors. 
Since manual detection of AgNORs is laborious, automation is of high interest. We present a deep learning-based pipeline for automatically determining the AgNOR-score from histopathological sections. An additional annotation experiment was conducted with six pathologists to provide an independent performance evaluation of our approach.
Across all raters and images, we found a mean squared error of 0.054 between the AgNOR-scores of the experts and those of the model, indicating that our approach offers performance comparable to humans.
\end{abstract}

\section{Introduction}
Nucleolar organizer regions (NORs) are nucleolar substructures involved in ribosomal RNA transcription \cite{ploton1986improvement}. NORs can be visualized in histological tissue sections using silver-based staining because of the silver affinity of two argyrophilic proteins involved in rRNA transcription and processing \cite{ploton1986improvement, DERENZINI2000117}. The silver-stained argyrophilic NORs are defined as AgNORs \cite{DERENZINI2000117}. The number of AgNORs per nuclei has been reported to be correlated with cell proliferation in vitro and the rate of tumor growth in vivo, as more malignant neoplasms tend to have more numerous and smaller AgNORs than benign or less malignant tumors \cite{webster_1,crocker1989should}. Furthermore, the AgNOR-score was reported to have significant prognostic value for outcome variables like survival \cite{kiupel1998prognostic, pich1994argyrophilic}. When examining the AgNOR-score in clinical practice, 100 randomly selected nuclei are examined per tissue section under a light microscope, while carefully focusing through the whole thickness of the slide \cite{kiupel1998prognostic}. The mean number of AgNORs per nucleus is then calculated from the assessed nuclei.
A limitation of the AgNOR-score is the considerable time effort for the examining pathologists. Deep learning methods are ideal here, as they can accelerate the process. However, there have been only a few attempts in the direction of deep learning based AgNOR scoring. One work in this area comes from Amorim et al., who studied the automatic assessment of AgNORs in cytological samples from cervical cancer using a U-Net architecture \cite{amorim2020novel}. In cytological samples, cells are well separated upon a bright background, however, in histological samples, cell borders are often obscured, with darker intervening stroma, rendering histological samples more difficult for automated AgNOR assessments. This work thus provides a first attempt at automatic scoring of histopathological samples for AgNOR using deep learning.
We argue that it is not the segmentation of individual AgNORs within a nucleus that is relevant to determine the AgNOR-score, but their quantity. Therefore, defining this task as a segmentation problem only introduces more unnecessary manual annotation effort for the annotation expert. To circumvent this, we define the problem as an object detection task where the number of AgNORs within a nucleus represents the label.
To quantify AgNORs, we used an anchor-free, state-of-the-art object detection algorithm. 
To evaluate the performance of our approach, we conducted an annotation experiment in which the AgNOR-scores for ten images, all representing individual tumor cases, were assessed by six pathology experts. For this study, we evaluated both AgNOR-scores per image and AgNOR-scores for individual nuclei.
The main contributions of this work are:
\begin{enumerate}
    \item We address the automatic assessment of AgNOR in histopathology using deep learning.
    \item We provide a baseline performance using an ensemble of state-of-the-art object detectors and a tailored sampling scheme.
    \item We perform extensive evaluations on the results of the human rater experiment.
\end{enumerate}

\section{Materials and methods}
\subsection{Data}
\begin{figure}[ht]
    \title{Figure}
    \includegraphics[scale=1]{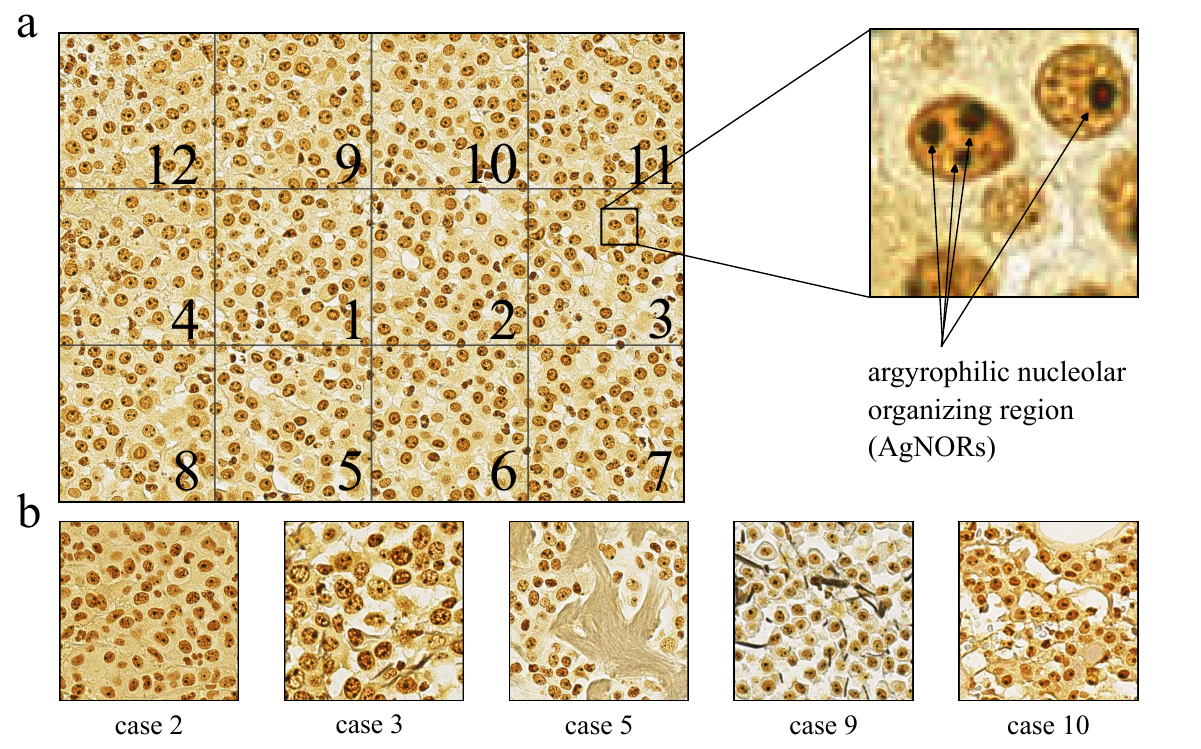}
    \caption{The left side of Figure a) shows a CCMCT section stained with silver-based staining like described in \cite{webster_1}. The grid used in the experiment is overlaid. On the right side of Figure a), a detailed representation of two nuclei with marked AgNORs is displayed. In Figure b), fields from randomly selected cases of the hold-out test set are shown. }
    \label{fig:fig2}
\end{figure}

In this work, we used 29 images ($1569 \times 1177$ pixel each) of canine cutaneous mast cell tumor (CCMCT). Each image shows a region of interest selected from a whole slide image by a pathologist. AgNOR histochemical staining was performed using a silver-based stain as described in \cite{webster_1}. The slides were digitized using an Aperio ScanScope CS2 scanner with a spatial resolution of 0.25 microns per pixel. All nuclei present in the images were annotated by a pathologist, with the number of identifiable AgNORs representing the label for each nucleus. This resulted in a total of 23,036 bounding box annotations grouped into twelve classes. The twelfth class includes all nuclei in which more than ten AgNORs are identifiable. The class distribution is strongly skewed toward classes with fewer than four AgNORs per nuclei, as can be seen in Tab. \ref{tab:1}. Of these images, ten images are used as a hold-out test set. An overview of samples from different cases of the test set is given in Fig.~\ref{fig:fig2}.

\begin{table}[ht]
\centering
\begin{tabular*}{\textwidth}{l@{\extracolsep\fill}|lllllllllllll}
number of AgNORs per nucleus & 0    & 1    & 2    & 3    & 4   & 5   & 6   & 7  & 8  & 9  & 10 & \textgreater 10 \\ \hline
number of cells  & 5124 & 9634 & 4620 & 1951 & 886 & 403 & 209 & 78 & 48 & 28 & 26 & 29 
\end{tabular*}
\caption{Number of instances in different classes.}
\label{tab:1}
\end{table}

\subsection{Human rater experiment}
To evaluate the algorithm's performance compared to human experts, we conducted a study with six pathology experts. The experts were asked to determine the number of AgNORs of at least one hundred nuclei for the ten images which are part of the hold-out test set. To reduce inter-rater discordance between the experts, the selection of nuclei to be annotated for the count of AgNORs was modified slightly from the clinical standard. The images used in the study were each divided into 12 fields, which were numbered like in Fig.~\ref{fig:fig2}. The experts were asked to process the fields one after another until at least 100 nuclei had been annotated. If the annotation was started in one field, the experts were instructed to annotate all remaining nuclei within this field, even if more than 100 nuclei were annotated in total. The AgNOR-scores for the corresponding image were derived by averaging the annotations of each expert.
The mean of the AgNOR-scores per case of all experts for an image was used as the expert value for this image, which is later compared to the algorithmically determined value.
To measure the agreement between the algorithm and the experts at the object level, we aggregated ground truth labels from the expert annotations. Since the areas annotated by the experts vary, we only aggregated class-level labels for the first field of the grid in Fig. \ref{fig:fig2}. A nucleus was taken as ground truth if it was annotated by at least two experts and assigned to the same class. If one nucleus was annotated by more than two experts, the class label was determined by a majority vote. To determine the inter-observer agreement, Fleiss's kappa coefficient and the intraclass correlation coefficient (ICC) were used \cite{landis1977measurement,shrout1979intraclass}. We report the ICC for a random set of $k$ raters where every instance is rated by all $k$ raters as described by Shrout and Fleiss \cite{shrout1979intraclass}.

\subsection{AgNOR-scoring pipeline}
To determine the AgNOR-score, we consider the problem an object detection task, where nuclei are discriminated based on the number of identifiable AgNORs. For detection, we used the fully convolutional one-stage object detector (FCOS) by Tian et al. \cite{tian2020fcos}, which is a state-of-the-art object detector that does not use anchor boxes like previous approaches and thus comes with a reduced set of hyperparameters to tune. To account for imbalances in the training data, we ensembled five detection models which were trained on five random subsets of the training data. Each of the five models produced its own AgNOR-score by averaging the class labels of all nuclei detected in one image. We used the median to aggregate the results of the ensemble of models per image. 
During training, we oversampled classes with lower support because of the class imbalance present in the training data. For the same reason, we used a small patch size of 128 $\times$ 128 pixels for training since this gave us better control over the classes present on each patch.
All models were trained until convergence as observed by the validation loss with a learning rate of $10^{-4}$ and Adam as optimizer. 
To ensure comparability between the results of the study and those of the algorithm, the inference of the detectors was performed analogously to the annotation of the experts described before. The respective image was divided into the same 12 fields, which needed to be processed in the same order as in the study until at least 100 nuclei were detected and classified.

\section{Results}
\begin{figure}[ht]
    \centering
    \includegraphics[scale = 1]{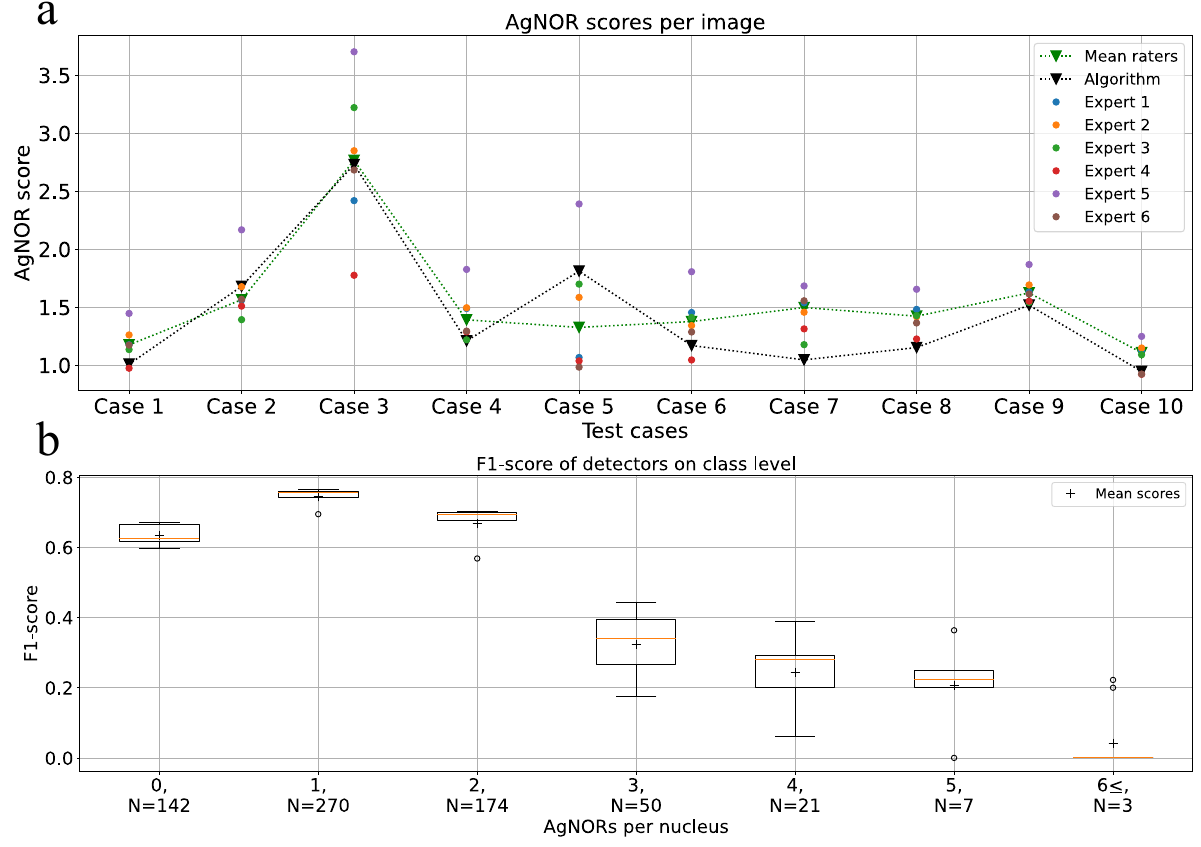}
    \caption{The AgNOR-scores of the different raters across the cases of the test data set are shown in Figure. a). The green triangles represent the image label formed from the mean of the AgNOR-scores of the pathologists. The black triangles depict the AgNOR-scores of the algorithm. Figure b) shows the performance of the individual object detectors of the ensemble at object level. The number of instances per class in the ground truth labels is specified by N.}
    \label{fig:main_fig}
\end{figure}

Figure \ref{fig:main_fig} shows the individual AgNOR-score for each rater, as well as the mean label for each case. 
The highest mean AgNOR-score across raters was 2.774 with a standard deviation of 0.500 and found for the third case in the set. The lowest mean AgNOR-score across raters was 1.075 with a standard deviation of 0.119. The average of these mean AgNOR-scores of all experts when calculated across all cases is 1.550 with a standard deviation of 0.441.
Between the nucleus-level annotations made by the experts, we found a Fleiss' kappa value of 0.545 and an ICC value of 0.390. According to the verbal description by Landis et al. \cite{landis1977measurement} the Fleiss' kappa value corresponds to a moderate agreement. 
In general, we found a slight disagreement between the AgNOR values determined by the algorithm and the experts' AgNOR-scores. The mean squared error between the algorithm-determined AgNOR-scores and those of the experts was 0.054, the mean absolute error was 0.201. 
Evaluating the model using the object level labels derived from the expert annotations, we found satisfactory performance in detecting classes of up to 2 AgNORs per nucleus, with mean F1-scores greater than 60\%. All classes with more than 2 AgNORs per nucleus were not detected as accurately, with mean F1 values of less than 40\%. Since there were only three nuclei within the ground truth with greater than or equal to six AgNORs, we combined the results from these classes into one box in Fig. \ref{fig:main_fig}.

\section{Discussion}
The results of the annotation experiment show that there are individual tendencies in assigning higher or lower AgNOR-scores among raters. The results of the experiment for case three suggest that the inter-rater concordance degrades with the average AgNOR-score per nucleus, and thus might coincide with proliferation speed.
A reason for this might be that more malignant neoplasms have more numerous and smaller AgNORs as previously described. Therefore, the AgNORs in case three might also be smaller, making it harder for the experts to distinguish them from each other and leading to a higher variance in the assessed AgNOR-scores.
The moderate Fleiss' kappa and ICC values show that the experts had difficulty arriving at a consistent result for different nuclei.
Since we performed the experiment with digital slides using a single $z$ plane, focusing through each cell individually, as stated in \cite{kiupel1998prognostic}, was not possible. One possible solution to mitigate this issue might be the use of stacks of images at multiple focus lengths. In this process, a slide is scanned in its entirety at various levels of focus and the multiple images are stitched together to form a three-dimensional image that allows for digital post-hoc focusing. Moreover, the use of higher scans with severely increased resolution could help experts to better separate individual AgNORs.
Overall, the low error between the algorithmically determined values and those of the experts shows that the algorithm is able to determine AgNOR-scores with human-like performance.
Its tendency to predict lower AgNOR-scores is presumably a consequence of the considerable class imbalance in the training data set. As a result, the algorithm confuses the higher classes, which is also reflected in the detection scores of the individual detectors in Fig. \ref{fig:main_fig}. Here, the detectors of the ensemble achieve favorable F1 scores on the three most dominant classes, whereas from three AgNORs per nucleus on, the performance decreases significantly. Nevertheless, this only has a minor effect on the estimation error on image-level due to the distribution of AgNORs per nucleus (see Tab. \ref{tab:1}). We encourage the use of fully automated pipelines such as this to investigate the utility and prognostic significance of AgNOR in future clinical tumor research. 

%

\printbibliography
\end{document}